\documentclass{article}

     \usepackage[nonatbib,preprint]{neurips_2020}

\usepackage[utf8]{inputenc} 
\usepackage[T1]{fontenc}    
\usepackage{hyperref}       
\usepackage{url}            
\usepackage{booktabs}       
\usepackage{amsfonts}       
\usepackage{nicefrac}       
\usepackage{microtype}      
\usepackage{graphicx}       
\usepackage{subcaption}     
\usepackage{adjustbox}

\title{TB-Net: A Tailored, Self-Attention Deep Convolutional Neural Network Design for Detection of Tuberculosis Cases from Chest X-ray Images}

\author{Alexander Wong$^{1,2,3}$, James Ren Hou Lee$^{3}$, Hadi Rahmat-Khah$^{3}$\\
\textbf{Ali Sabri$^{4}$ and Amer Alaref$^{5,6}$}\\
$^1$Department of Systems Design Engineering, University of Waterloo, Canada\\
$^2$Waterloo Artificial Intelligence Institute, Canada\\
$^3$DarwinAI Corp., Canada\\
$^4$Department of Radiology, Niagara Health, McMaster University, Canada\\
$^5$Department of Diagnostic Radiology, Thunder Bay Regional Health Sciences Centre, Canada\\
$^6$Department of Diagnostic Imaging, Northern Ontario School of Medicine, Canada\\
}

\begin{document}

\maketitle

\begin{abstract}
Tuberculosis (TB) remains a global health problem, and is the leading cause of death from an infectious disease. A crucial step in the treatment of tuberculosis is screening high risk populations and the early detection of the disease, with chest x-ray (CXR) imaging being the most widely-used imaging modality. As such, there has been significant recent interest in artificial intelligence-based TB screening solutions for use in resource-limited scenarios where there is a lack of trained healthcare workers with expertise in CXR interpretation. Motivated by this pressing need and the recent recommendation by the World Health Organization (WHO) for the use of computer-aided diagnosis of TB in place of a human reader, we introduce TB-Net, a self-attention deep convolutional neural network tailored for TB case screening. More specifically, we leveraged machine-driven design exploration to build a highly customized deep neural network architecture with attention condensers. We conducted an explainability-driven performance validation process to validate TB-Net's decision-making behaviour. Experiments on CXR data from a multi-national patient cohort showed that the proposed TB-Net is able to achieve accuracy/sensitivity/specificity of 99.86\%/100.0\%/99.71\%. Radiologist validation was conducted on select cases by two board-certified radiologists with over 10 and 19 years of experience, respectively, and showed consistency between radiologist interpretation and critical factors leveraged by TB-Net for TB case detection for the case where radiologists identified anomalies.  While not a production-ready solution, we hope that the open-source release of TB-Net\footnote{\url{https://github.com/darwinai/TuberculosisNet}} as part of the COVID-Net initiative will support researchers, clinicians, and citizen data scientists in advancing this field in the fight against this global public health crisis.

\end{abstract}

\section{Introduction}

Tuberculosis (TB) remains a devastating global public health crisis, with tremendous on-going negative impact around the world as the leading cause of death for an infectious disease at approximately 1.5 million deaths and 10 million infected each year~\cite{who2020}.  Caused
by Mycobacterium tuberculosis (M. tb), a species of pathogenic bacteria, TB is an airborne disease that affects approximately a quarter of the global population, particular areas of the world faced by poverty and economic distress~\cite{who2020}.  More specifically, the most devastating effect of tuberculosis has been on low- and middle-income regions, with two thirds of all tuberculosis infections found in 8 countries (Bangladesh, China, India, Indonesia, Nigeria, Pakistan, Philippines, and South Africa)~\cite{who2020}.  Tuberculous is a curable disease, with approximately 85\% of infections successfully treated via a 6-month drug treatment regimen of different antibiotics.

A very crucial step in the treatment of tuberculosis is screening high risk populations and the early detection of the disease, although tuberculous remains underdiagnosed with an estimated 3 million out of 10 million infected individuals not diagnosed or reported to the World Health Organization (WHO)~\cite{whoscreening2020}.  Currently, the most widely-used imaging modality used in the screening of tuberculous is chest x-ray (CXR) imaging, which has been shown to be a highly effective and cost-effective screening tool in this scenario~\cite{Hoog,Diaz,Yip}. However, one of the biggest limitations with the use of CXR for tuberculosis screening is that it requires experienced human readers such as radiologists or trained clinicians and technicians for interpretation~\cite{who2020}, especially given the significant shortage of such experienced readers in the most effected countries.

Given the shortage of experienced readers worldwide for CXR interpretation for tuberculosis screening, there has been significant recent interest in artificial intelligence-based TB screening solutions for use in resource-limited scenarios~\cite{Melendez,Singh,Hooda,Hernandez,Yadav,Ahsan,Meraj,Lopes,Nguyen,Rahman,Vajda,Lakhani,Chauhan,Hwang,Pasa}.  In fact, the most recent WHO guidelines for tuberculosis screening introduced a new recommendation stating that, for those 15 years and older in populations where screening is recommended, computer-aided detection (CAD) may be used in place of human readers for CXR interpretation for tuberculosis screening~\cite{whoscreening2020}.  Most recently, a comprehensive study conducted by Rahman et al.~\cite{Rahman} compared nine different deep convolutional neural network architectures (ResNet-18~\cite{resnet}, ResNet-50, ResNet-101, ChexNet~\cite{rajpurkar2017chexnet}, Inception-V3~\cite{szegedy2015rethinking}, VGG-19~\cite{simonyan2015deep}, DenseNet-201~\cite{huang2018densely}, SqueezeNet~\cite{iandola2016squeezenet}, and MobileNet-v2~\cite{sandler2019mobilenetv2}) for the task of detecting TB patient cases from CXR images, and found the ChexNet deep neural network architecture design~\cite{rajpurkar2017chexnet}, a state-of-the-art deep neural network architecture for CXR image analysis, to provide the highest sensitivity and specificity without the use of segmentations.

Motivated by this pressing need and the new recommendation by the WHO for the use of CAD for tuberculosis screening, we introduce TB-Net, a self-attention deep convolutional neural network design tailored for TB case screening. More specifically, we leveraged machine-driven design exploration to build a highly customized deep neural network architecture with attention condensers. The TB-Net deep neural network design is not only designed to be high-performing but also highly efficient, which is especially important for practical, operational TB screening given that high-risk regions for TB around the world are those faced by poverty and economic distress, and as such have high resource constraints.  We conducted an explainability-driven performance validation process to validate TB-Net's decision-making behaviour. Furthermore, radiologist validation was conducted with two board-certified radiologists to study the consistency between radiology interpretation and TB-Net's decision-making behaviour.

The proposed TB-Net is part of the COVID-Net open source initiative\footnote{\url{http://www.covid-net.ml}}~\cite{covidnet,alex2020covidnets,Gunraj2020,gunraj2021covidnet,ebadi2021covidxus}, which was launched to accelerate advancements in machine learning for tackling different challenges ranging from screening to risk stratification to treatment planning for patients with the severe acute respiratory syndrome coronavirus 2 (SARS-CoV-2).  Accordingly to the WHO, it is anticipated that those with both tuberculosis and SARS-CoV-2 infections could potentially experience poorer treatment outcomes, particularly if tuberculosis treatment is interrupted as a result of SARS-CoV-2 infection~\cite{who3}.  Furthermore, tuberculosis and SARS-CoV-2 infection can share similar symptoms~\cite{who3}.  Therefore, effective CAD of both SARS-CoV-2 and tuberculosis infections to support clinicians and front-line healthcare workers can have great potential for improving clinical workflows for tackling these health crises by improving screening, triaging, risk stratification, and treatment planning.

While not a production-ready solution, we hope that the open-source release of TB-Net will support researchers, clinicians, and citizen data scientists in advancing this field in the fight against this global public health crisis.

\begin{figure*}[!t]
  \centering
  \includegraphics[width= \linewidth]{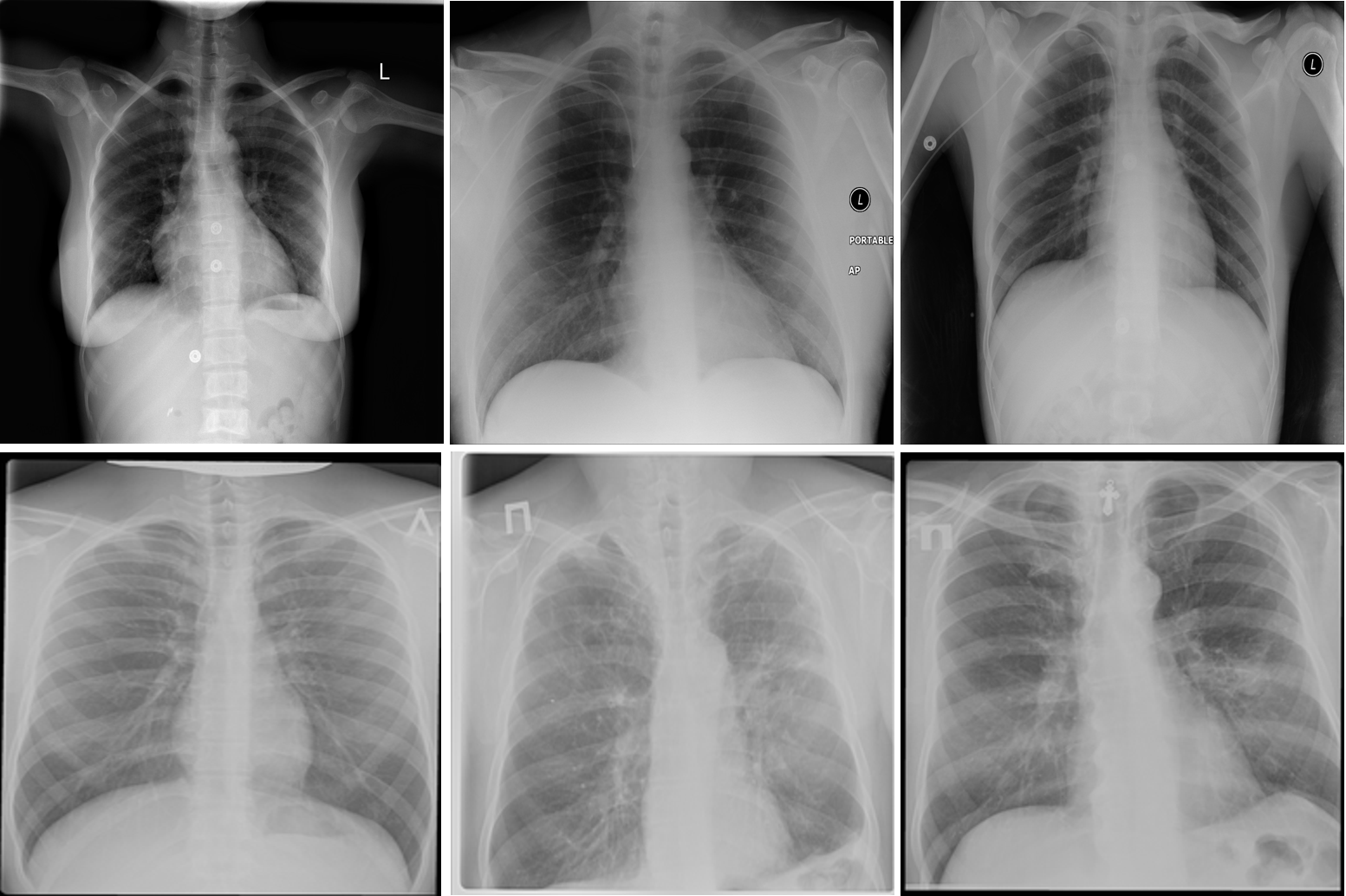}
  \caption{Example chest X-ray images from the multi-national patient cohort introduced by Rahman et al.~\cite{Rahman}: (top) TB negative patient cases and (bottom) TB positive patient cases.}
  \label{fig:cxr-examples}
\end{figure*}

The paper is organized as follows. Section 2 describes the underlying methodology behind the design of the proposed TB-Net, data preparation, explainability-driven performance validation, and radiologist validation.  Section 3 presents and discusses the efficacy and decision-making behaviour of TB-Net both quantitatively and qualitatively, along with radiologist validation.  Finally, conclusions are drawn and discussions in broader impact and future directions are presented in Section 4.

\section{Methods}

In this study, we introduce TB-Net, a self-attention deep convolutional neural network design for detection of TB cases from chest X-ray images.  A machine-driven design exploration strategy is leveraged to automatically discover highly customized and unique macro-architecture and micro-architecture designs that make up the proposed TB-Net self-attention deep neural network architecture with attention condensers.  Explainability-driven performance validation was conducted to study and validate the decision-making behaviour of TB-Net.  Finally, radiologist validation was conducted by two board-certified radiologists.  The details between data preparation, network design, explainability-driven performance validation, and radiologist validation are described below.

\subsection{Data preparation}
To train and evaluate the proposed TB-Net, we leveraged the CXR data from a multi-national patient cohort introduced in a study by Rahman et al.~\cite{Rahman}, which unified patient cohorts from several initiatives from around the world~\cite{Jaeger,Belarus,NIAID,RSNA}.  More specifically, the multi-national patient cohort consists of patient cohorts curated by the Department of Health and Human Services in Montgomery County, Maryland, USA, Shenzhen No. 3 People’s Hospital in China, the National Institute of Allergy and Infectious Diseases in the USA, as well as the Radiological Society of North America. This multi-national patient cohort represents one of the largest, most diverse patient cohorts for exploring computer-aided tuberculosis screening, to the best of the authors' knowledge.

After additional image quality screening of the CXR images, the CXR data used in this study comprises 6,939 CXR images. In terms of data distribution, there are a total of 3,461 CXR images from TB positive patients and 3,478 CXR images from TB negative patients.  The training, validation, and test data consist of 80\%, 10\%, and 10\% of the patient cases randomly selected from the multi-national patient cohort, respectively.  To facilitate for the training and evaluation of TB-Net, the CXR images were resampled to 224$\times$224 and mean imputation was performed on the top left-hand and top right-hand corners to mitigate the presence of embedded markings found in the CXR images.  Example CXR images from the multi-national patient cohort used in this study for both TB negative and TB positive patient cases are shown in Figure~\ref{fig:cxr-examples}.

All data generation and preparation scripts are available in an open source manner at \url{https://github.com/darwinai/TuberculosisNet}.

\begin{figure*}[t]
  \centering
  \includegraphics[width= \linewidth]{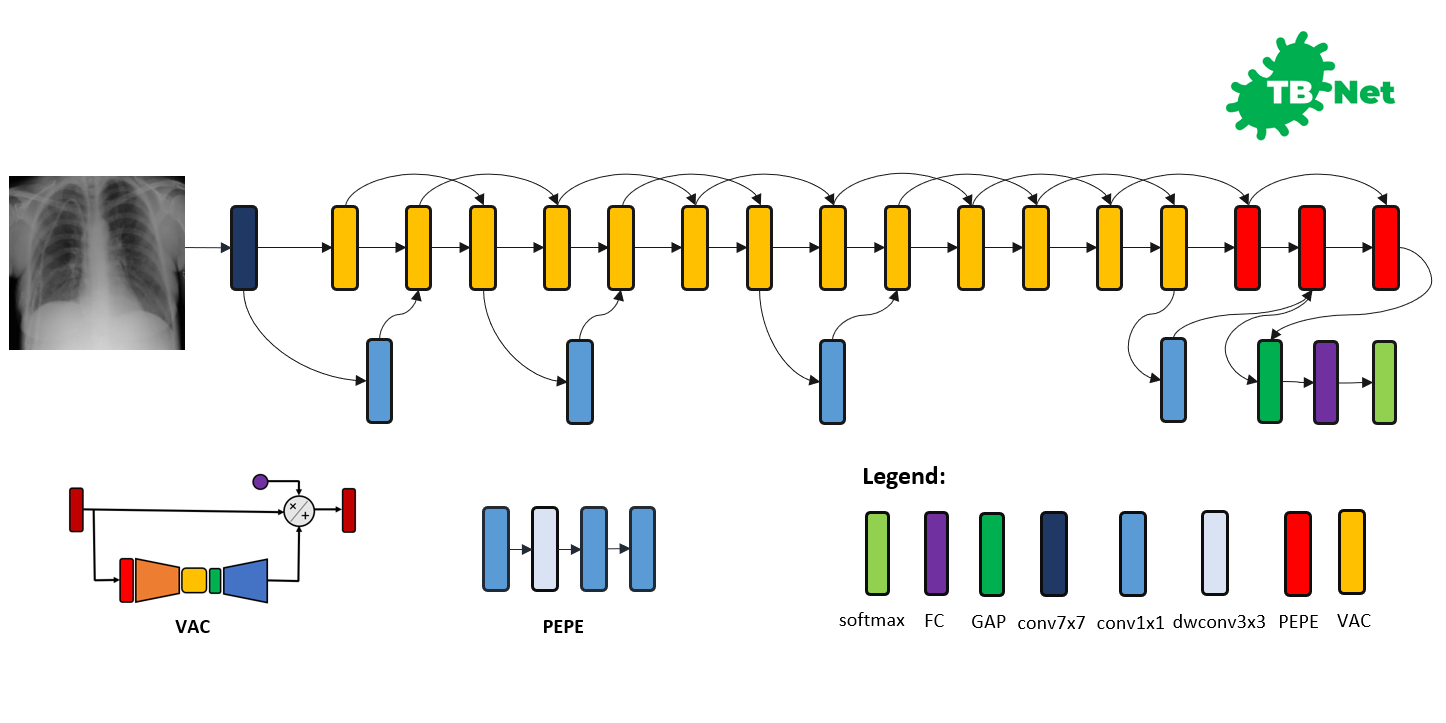}
  \caption{The proposed TB-Net architecture design.  The TB-Net design exhibits high architectural heterogeneity, light-weight design patterns, and the utilization of visual attention condensers, with macro-architecture and micro-architecture designs tailored specifically for the detection of TB cases from chest X-ray images.}
  \label{fig:architecture}
\end{figure*}

\subsection{Network design}
The proposed TB-Net self-attention deep neural network architecture design was constructed using a machine-driven design exploration strategy using the aforementioned CXR data from the multi-national patient cohort.  More specifically, we leverage the concept of generative synthesis~\cite{gensynth} to determine the macro-architecture and micro-architecture designs of a deep neural network architecture tailored for the task of TB case detection from CXR images.  The designs are discovered automatically using an optimal generator $\mathcal{G}$ that is capable of, given a set of seeds $S$, generating deep neural network architectures $\left\{N_s|s \in S\right\}$ that maximize a universal performance function $\mathcal{U}$ (e.g., \cite{wong2018netscore}) under a set of constraints defined by an indicator function $1_r(\cdot)$,

\begin{equation}
\mathcal{G}  = \max_{\mathcal{G}}~\mathcal{U}(\mathcal{G}(s))~~\textrm{subject~to}~~1_r(\mathcal{G}(s))=1,~~\forall s \in S.
\label{eqoptimization}
\end{equation}

\noindent The set of constraints for building TB-Net as defined via indicator function $1_r(\cdot)$ were: (1) sensitivity $\geq$ 95\%, (2) specificity $\geq$ 95\%, and (3) number of parameters $\leq$ 5M.

The proposed TB-Net self-attention deep convolutional neural network design is shown in Figure~\ref{fig:architecture}. A number of interesting observations can be made.  First, it can be observed that the overall network architecture design exhibits high macro-architecture and micro-architecture heterogeneity, with a mix of standard convolutions, depth-wise convolutions, point-wise convolutions, and self-attention mechanisms with different micro-architecture characteristics.  This high degree of architectural diversity and heterogeneity reflects the fact that a machine-driven design exploration strategy was leveraged to customize the design in a very fine-grained manner specifically around TB case detection using CXR images to achieve an optimal level of performance for the given task at hand.

Second, it can be observed that the network architecture possesses a very light-weight design, consisting primarily of highly efficient depth-wise convolutions and point-wise convolutions as well as leveraging light-weight design patterns such as project-expansion-projection-expansion (PEPE) patterns that reduces and expands representational dimensionality in a way that strikes an optimal balance between accuracy and efficiency.  This light-weight design pattern utilization reflects the ability for a machine-driven design exploration strategy to tailor the macro-architecture design of deep neural network architectures based on the architectural complexity constraint imposed in the indicator function.  The highly efficient architecture design of TB-Net is especially critical to enabling potential widespread adoption since real-world TB screening scenarios in high-risk regions faced by poverty and economic distress have high resource and cost constraints, and so deployment would often have to take place on low-cost, low-end computing devices.

Third, it can be observed that a majority of the layers of the TB-Net network architecture design is comprised of visual attention condensers~\cite{wong2020attendnets}, which are a variant of the highly efficient attention condenser self-attention mechanisms recently introduced in~\cite{wong2020tinyspeech}.  More specifically, visual attention condensers produce condensed
embedding characterizing joint spatial and cross-channel activation relationships and achieves selective attention accordingly to improve representational capability while maintaining very low architectural and computational complexity.  As a result, by leveraging visual attention condensers, the proposed TB-Net network architecture design facilitates for high TB screening performance by better focusing its attention on the distinguishing visual cues within CXR images for identifying TB positive patient cases in a very efficient manner.

Last but not least, the end-stage sub-architecture of the TB-Net deep neural network architecture consists of a global average pooling layer, a fully-connected layer, and a softmax layer to produce the final output for predicting whether a patient is TB positive or negative.  The TB-Net network is available in an open source manner at \url{https://github.com/darwinai/TuberculosisNet}.

\subsection{Network training}

Training was conducted on the proposed TB-Net deep neural network architecture design using stochastic gradient descent optimization with a learning rate of 0.0001, momentum of 0.9, and a batch size of 8 for 200 epochs.  Data augmentation was conducted during training with the following augmentation types: horizontal flip, random cropping (within 10\%), random contrast shift (within 20\%), and random intensity shift (within 10\%). All construction, training, and evaluation are conducted in the TensorFlow deep learning framework.

The scripts for the aforementioned process are available in an open source manner at \url{https://github.com/darwinai/TuberculosisNet}.

\subsection{Explainability-driven performance validation}
To gain a deeper insight and validate the decision-making behaviour of TB-Net, we leveraged GSInquire~\cite{gsinquire}, a state-of-the-art explainability method that was shown to no only provide explanations that better reflect the decision-making behaviour of deep neural networks when compared to other well-known methods, but also identify specific critical factors that are quantitatively critical to the decision-making process rather than relative heatmaps pertaining to relative importance variations.

Briefly, GSInquire leverages an inquisitor $\mathcal{I}$ within a generator-inquisitor pair $\left\{\mathcal{G},\mathcal{I}\right\}$ during the generative synthesis~\cite{gensynth} process used in the machine-driven exploration strategy.  The inquisitor can be more formally defined as $\mathcal{I}(\mathcal{G};\theta_\mathcal{I})$ parameterized by $\theta_\mathcal{I}$ that, given a generator $\mathcal{G}$, produces  $\Delta\theta_\mathcal{G}$ (i.e., $\Delta\theta_\mathcal{G} = \mathcal{I}(\mathcal{G})$).  To obtain an interpretation $z$ of a decision from a network $N=\mathcal{G(s)}$ comprising a set $V$ of vertices $v \in V$ and a set $E$ of edges $e \in E$ (here, the TB-Net network) for an input signal $x$ (here, a CXR image), the inquisitor $\mathcal{I}$ probes $\left\{\mathcal{V}_{s},\mathcal{E}_{s}\right\}$, where
$\mathcal{V}_{s} \subseteq V_{s}$ and $\mathcal{E}_{s} \subseteq E_{s}$ with targeted stimulus $x$,
and the resulting set $Y_{\mathcal{G}\left(s\right)}$ of reactionary responses
$y \in Y_{\mathcal{G}\left(s\right)}$ are observed and used to update $I$.  After the update of $I$, $\Delta\theta_\mathcal{G} = \mathcal{I}(\mathcal{G})$ is generated, transformed, and projected into same subspace as $x$ via a transformation  $\mathcal{T}(\Delta\theta_{\mathcal{G}\left(s\right)})$ to create an interpretation $z(x;N)$.  The details related to the use of GSInquire to generate interpretations of deep neural network decision-making behaviour for CXR images can be found in~\cite{covidnet}.  Here, the interpretation $z$ indicates the critical factors leveraged by TB-Net in its decision-making process for a CXR image.

Explainability-driven performance validation facilitates for: 1) transparent validation of the TB-Net network to ensure that the TB case detection process is primarily driven by clinical relevant visual indicators such as infiltrates, consolidations, pleural effusion, cavities, and lesions, 2) identification of potentially erroneous visual indicators being leveraged such as embedded markers and text, imaging artifacts, and motion artifacts, and 3) improve greater trust in the clinical workflow through greater transparency.

\subsection{Radiologist validation}
The results for TB-Net that were obtained during the explainability-driven performance validation process for select patient cases are further reviewed and reported on by two board-certified
radiologists (A.S. and A.A.). The first radiologist (A.S.) has over 10 years of experience, and the second radiologist (A.A.) has over 19 years of radiology experience.

\section{Results and Discussion}

We evaluate the efficacy of the proposed TB-Net self attention deep convolutional neural network design for detecting TB cases from CXR images in three ways.  First, we evaluate the quantitative performance of the network as well as study its architectural and computational complexity.  Second, we study its decision-making behaviour using an explainability-driven performance validation strategy.  Third, we conduct radiologist validation on study the consistency of TB-Net's decision-making behaviour with radiologist interpretation.  The details of the quantitative and qualitative results are discussed below.

\begin{table}[!t]
\caption{Accuracy, sensitivity, and specificity of TB-Net on the test data from the multi-national patient cohort. Better performance metric in \textbf{bold}.}
\begin{center}
\begin{tabular}{|c|c|c|c|c|}
\hline
      \textbf{Architecture} & \textbf{Accuracy ($\%$)} & \textbf{Sensitivity ($\%$)} & \textbf{Specificity ($\%$)} \\
      \hline
       CheXNet~\cite{rajpurkar2017chexnet} & 99.42 & \textbf{100} & 98.85  \\
\hline
 TB-Net & \textbf{99.86} & \textbf{100} & \textbf{99.70} \\
\hline
\end{tabular}\par
\label{tab:quant_results}
\end{center}
\end{table}

\subsection{Quantitative analysis}

The accuracy, sensitivity, and specificity of the proposed TB-Net are shown in Table~\ref{tab:quant_results}, while the architectural and computational complexity of the proposed TB-Net are shown in Table~\ref{tab:quant_results2}.  For comparison purposes, CheXNet~\cite{rajpurkar2017chexnet} was also evaluated, given that it is a state-of-the-art deep neural network architecture for CXR image analysis and was found in a comprehensive study conducted by Rahman et al.~\cite{Rahman} to be the best performing deep neural network architecture amongst nine different deep convolutional neural network architectures (ResNet-18~\cite{resnet}, ResNet-50, ResNet-101, ChexNet~\cite{rajpurkar2017chexnet}, Inception-V3~\cite{szegedy2015rethinking}, VGG-19~\cite{simonyan2015deep}, DenseNet-201~\cite{huang2018densely}, SqueezeNet~\cite{iandola2016squeezenet}, and MobileNet-v2~\cite{sandler2019mobilenetv2}) for the task of detecting TB patient cases from CXR images without the use of segmentations.

A number of observations can be made based on the quantitative performance results.  First, it can be observed from Table~\ref{tab:quant_results} that the TB-Net network achieved high accuracy, sensitivity, and specificity of 99.86\%, 100\%, and 99.7\%, respectively, and thus achieves the same level of sensitivity and slightly higher specificity when compared to CheXNet used in this study.  The high sensitivity achieved with the proposed TB-Net implies that there would be fewer missed TB positive patients during the TB screening process, which is highly desirable from a clinical perspective especially given the infectious nature of TB and the need to reduce spread within the community.  On the other hand, the high specificity achieved with the proposed TB-Net implies that there would be fewer false positive detections, which is important to reduce the burden on healthcare systems caused by additional work for clinicians and front-line healthcare workers.

Second, it can be observed from Table~\ref{tab:quant_results2} that the TB-Net network achieves low architectural complexity and computational complexity of 4.24 million parameters and 0.42 billion multiply-accumulate (MAC) operations, which is also $\sim$1.9$\times$ lower and $\sim$6.7$\times$ lower, respectively, than that of CheXNet used in this study.  The high architectural and computational efficiency achieved by the proposed TB-Net network is important for enabling CAD-driven TB screening on low-cost, low-power computing devices, particularly given the types of resource-constrained clinical environments faced in high-risk regions faced by poverty and economic distress.

These quantitative results illustrate that the self-attention network architecture design tailored via a machine-driven design exploration approach is capable of achieving high detection performance while maintaining high efficiency, thus illustrating high potential for resource-constrained clinical environments.

\begin{table}[!t]
\caption{Architectural complexity and computational complexity of TB-Net on the test data from the multi-national patient cohort. Better performance metric in \textbf{bold}.}
\begin{center}
\begin{tabular}{|c|c|c|}
\hline
      \textbf{Architecture} & \textbf{Params (M)} & \textbf{MACs (G)} \\
      \hline
       CheXNet~\cite{rajpurkar2017chexnet} & 8.06 & 2.8 \\
\hline
 TB-Net & \textbf{4.24} & \textbf{0.42} \\
\hline
\end{tabular}\par
\label{tab:quant_results2}
\end{center}
\end{table}

\subsection{Qualitative analysis}

\begin{figure*}[!t]
  \centering
  \includegraphics[width= \linewidth]{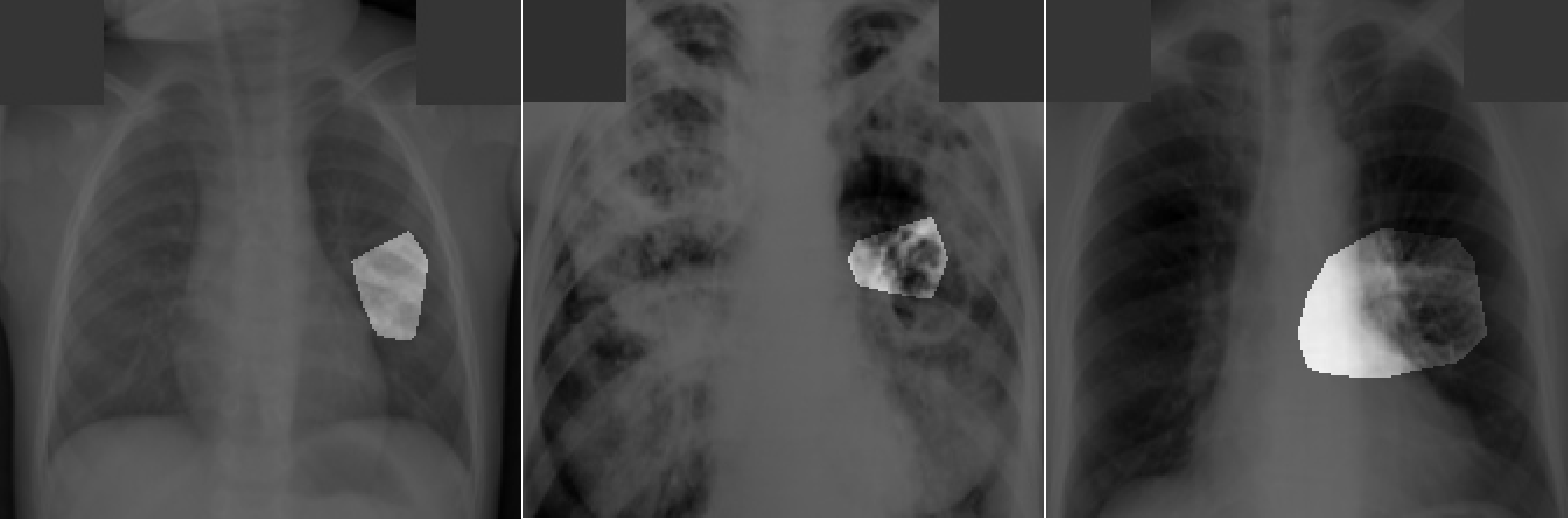}
  \caption{Examples of patient cases with associated critical factors (highlighted regions) as identified by GSInquire~\cite{gsinquire} during explainability-driven performance validation.  From left to right: (a) Case 1, (b) Case 2, and (c) Case 3.  Radiologist validation showed that several of the critical factors identified are consistent with radiologist interpretation.}
  \label{fig:explainability}
\end{figure*}

Explainabity-driven performance validation was conducted on TB-Net and  examples of patient cases with associated critical factors identified by GSInquire for driving the decision-making behaviour of the proposed TB-Net are shown in Figure~\ref{fig:explainability}. It can be seen that the proposed TB-Net is primarily relying on clinically relevant areas of the lung in the CXR images to drive its decision-making behaviour.  Furthermore, it can be seen that it is not relying on erroneous visual indicators such as motion artifacts, embedded symbols and text, and imaging artifacts.  As such, it can be seen that TB-Net is exhibiting clinically relevant decision-making behaviour.

\subsection{Radiologist analysis}
The expert radiologist findings for select patient cases in regards to the relevancy of the critical factors identified during the explainability-driven performance validation process as shown in Figure~\ref{fig:explainability} are as follows. In all three cases, TB-Net correctly detected them to be TB positive cases.

\textbf{Case 1.} According to radiologist findings, both radiologists did not observe any abnormalities that would be indicative of TB, while identified critical factors leveraged by TB-Net indicate some form of abnormality in the right midlung region.

\textbf{Case 2.} According to radiologist findings, it was observed by both radiologists that there is a cavity in the hilar region that coincide with the identified critical factors leveraged by TB-Net in that region.  One of the radiologists also observed other scattered opacities in the lungs.

\textbf{Case 3.} According to radiologist findings, both radiologists did not observe any abnormalities that would be indicative of TB, while identified critical factors leveraged by TB-Net indicate some form of abnormality in the right lower lung region.

As such, based on the radiologist findings and observations on the three patient cases, it was shown that the critical factors identified by GSInquire as critical factors driving the decision-making behaviour of TB-Net was consistent with radiologist interpretation for Case 2, but not all regions of concern as identified by the radiologists are necessarily leveraged by TB-Net in making its TB case detection decisions.  Furthermore, more interestingly, the critical factors identified by GSInquire as critical factors driving the decision-making behaviour of TB-Net for Case 1 and Case 3 for correctly determining the patients as TB positive were not identified by the radiologists.

\section{Conclusion and Discussion}
In this study, we introduced TB-Net, a self-attention deep convolutional neural network tailored for tuberculosis case screening. A machine-driven design exploration strategy was leveraged to build a highly customized deep neural network architecture with attention condensers. An explainability-driven performance validation process was conducted to validate TB-Net's decision-making behaviour, and was further confirmed via radiologist validation.   Experimental results demonstrate that TB-Net can not only achieve high tuberculosis case detection performance in terms of sensitivity and specificity, but also exhibit clinically relevant behaviour during an explainability-driven performance validation process as well as during the radiologist validation process for the case where radiologists identified anomalies.

Since tuberculosis is an on-going global health crisis and is curable if detected, the hope is that research such as TB-Net and open source initiatives such as the COVID-Net initiative that TB-Net is part of can accelerate the advancement and adoption of deep learning-driven computer aided diagnosis solutions within a clinical setting to aid front-line health workers and healthcare systems in improving clinical workflow efficiency and effectiveness in the fight against the on-going tuberculosis crisis in high-risk regions where there is a tremendous scarcity of experienced human readers for tuberculosis screening.  Therefore, additional care was taken to perform explainability-driven performance validation as well as radiologist validation to conduct additional checks and balance around the decision-making behaviour of TB-Net in a transparent and responsible manner.

To the best of the authors' knowledge, this research on tuberculosis screening using deep learning does not put anyone at any potential disadvantages.  However, it is important to note that TB-Net is not a production-ready solution and the current focus is on facilitating research advancements in the area.  While not a production-ready solution, the hope is that the open-source release of TB-Net as part of the COVID-Net initiative will support researchers, clinicians, and citizen data scientists in advancing this field in the fight against this global public health crisis.  Further work involves the exploration of tailored deep neural network designs for other tasks in the tuberculosis clinical workflow (e.g. severity assessment and treatment planning).

\section*{Author contributions statement}
    A.W. conceived the experiments, J.L. and H.R. conducted the experiments, all authors analysed the results, A.A. and A.S. reviewed and reported on select patient cases and corresponding explainability results illustrating model's decision-making behaviour, and all authors reviewed the manuscript.

\section*{Additional information}
    \textbf{Competing interests} A.W., J.L., and H.R. are affiliated with DarwinAI Corp.

\bibliographystyle{IEEEtran}
\bibliography{references}

\end{document}